# Corrugation induced stacking solitons with topologically confined states in gapped bilayer graphene

Si-Yu Li[1], Hua Jiang[2,*], Jiao-Jiao Zhou[2], Haiwen Liu[1], Fan Zhang[3], and Lin He[1,†]

[1]Center for Advanced Quantum Studies, Department of Physics, Beijing Normal University, Beijing, 100875, People's Republic of China

[2]College of Physics, Optoelectronics and Energy, Soochow University, Suzhou, 215006, People's Republic of China

[3]Department of Physics, University of Texas at Dallas, Richardson, Texas 75080, USA

Correspondence and requests for materials should be addressed to H.J. (e-mail: jianghuaphy@suda.edu.cn) and L.H. (e-mail: helin@bnu.edu.cn).

**Graphene, as an atomic-thick ultrasoft membrane, almost has no resistance against out-of-plane deformations and, therefore, it is always wrinkled to a certain degree. Recently, corrugated structures and their effects on the electronic properties of monolayer graphene have been studied extensively. However, similar experimental studies in bilayer graphene have yet to be reported. Here, we show that corrugations in bilayer graphene can generate incommensurate stacking solitons (domain walls) between commensurate Bernal-stacked domains. By using scanning tunneling microscopy, we microscopically study electronic structures of a corrugation-induced stacking soliton that separates two adjacent AB- and BA-stacked bilayer regions with a uniform interlayer potential difference. Both topological gapless edge states and quasi-localized gapped quantum-well-like states are observed in the nanoscale corrugation. Atomic resolution mapping of the topological edge states along the stacking soliton reveals the existence of intervalley scattering between them, which could explain recent experiments where the conductance along domain walls of gapped graphene bilayer is smaller than a quantized value $4e^2/h$ (here $e$ is the unit charge and $h$ is Planck's constant).**

The resistance to bend a graphene sheet only involves the π-orbital misalignment between adjacent pairs of the carbon atoms, which generates negligible resistance against out-of-plane deformations. Consequently, corrugations exist naturally and almost ubiquitously in graphene monolayer[1-6]. Recently, it has been demonstrated that the corrugated structures of graphene monolayer under certain circumstances can affect its electronic properties dramatically[7-13]. For example, lattice deformation in corrugated graphene can generate pseudomagnetic fields acting on the low-energy quasiparticles and result in pseudo-Landau quantization in the absence of magnetic fields[8-13]. In bilayer graphene, corrugations may lead to even more exotic phenomena because lattice deformation in corrugated structure of the bilayer graphene can further induce stacking faults. However, a systematic experimental study of corrugations in the bilayer graphene has been lacking to date. Here we use scanning tunneling microscopy (STM) to image atomic structures of corrugations in bilayer graphene grown on Rh foils, and demonstrate, for the first time, that a corrugation can generate an incommensurate stacking soliton, hosting valley-protected topological gapless modes[14-22], between two adjacent commensurate Bernal domains with AB and BA stacking orders, respectively.

The bilayer graphene studied in the current work was synthesized on Rh foils by using a facile ambient pressure chemical vapor deposition (CVD) method (see Methods and Fig. S1 of Supplemental Material for more details)[10,13]. The graphene grown on Rh foils is controlled by a segregation mechanism in which its thickness can be tuned by simply varying the cooling rate during the growth process[10,13]. Figure 1a shows Raman spectroscopy mapping of the as-grown graphene on a Rh foil. The regions with different colors reflect changes in thickness (layer number) of the graphene sample (see Fig. S2 of Supplemental Material for Raman spectra of graphene systems with different thickness)[23], and only the sample mainly covered with bilayer graphene is studied in this work. To identify the Bernal bilayer regions, we further employ both the STM images (Figure 1) and the scanning tunneling spectroscopy (STS) measurements (Figure 2). First, the atomic resolution STM image of the Bernal bilayer shows a triangular lattice, which is different from the hexagonal lattice of monolayer graphene. Second, the substrate breaks the inversion symmetry of the Bernal bilayer and produces a band gap of the order of 100 meV in the Bernal bilayer[22,24-26]. Such a large gap can be detected in STS measurements and help us to further identity the Bernal bilayer graphene.

Figure 1b shows a representative STM image of a Bernal bilayer region on the Rh foil. Obviously, there are many one-dimensional (1D) quasiperiodic ripples in bilayer graphene, as observed previously in monolayer graphene[6,11,13]. Mismatch of thermal expansion coefficients between graphene and the supporting substrates generates in-plane strain in graphene and results in the corrugated structures. In bilayer graphene, the in-plane strain not only induces lattice deformation in the corrugated structures, but may also lead to relative displacement of the adjacent bilayer, which consequently changes the local stacking order of the bilayer. Figure 1c shows a typical 1D nanoscale ripple where the stacking order of the bilayer graphene changes across it. In both domains left and right to the ripple, we observe triangular lattices reflecting the A/B sublattice asymmetry due to the inter-layer tunneling. By sharp contrast, in the central ripple region, we observe a hexagonal lattice (See Figure S3 of Supplemental Material for more STM images). This indicates the fact that the ripple in Fig. 1c generates a 1D incommensurate stacking soliton that separates two adjacent commensurate Bernal-stacked bilayer domains. By carefully examining the atomic structure around the ripple, we demonstrate that the left and right domains are respectively AB- and BA-stacked (see Figure S4 of Supplemental Material), as schematic shown in Fig. 1d. This suggests that the two adjacent graphene sheets have translated relative to each other in opposite directions across the ripple. Here we should point out that the stacking soliton studied in this work is different from the recently reported AB-BA domain wall[14,22,27], which exists naturally in bilayer graphene and is hard to generate in a designed way. However, the corrugation-induced stacking solitons can be generated and tuned by external stress. This provides a more controllable method to generate the stacking domain walls, which could be crucial for future studies of the large-scale applications and the interactions of multiple domain walls.

To investigate the influence of the stacking order and corrugation on the electronic spectra of bilayer graphene, we measure the spatial evolution of the tunneling spectra along a line across the ripple, as shown in Fig. 2a. In both the AB- and BA-stacked domains, we clearly observe a finite gap ∼ 170 meV induced by an effective electric field generated by the Rh foil. The observed bandgap agrees well with the range of values reported previously for Bernal bilayer on various substrates[19,24-26,28,29]. The spectra recorded in the ripple exhibit two notable features that are not observed in the two adjacent Bernal bilayer domains (see Figure S5 of Supplemental Material). One feature is the emergence of several peaks in the tunneling spectra. Similar peaks arising from strain-induced quasi-localized states have been observed previously in corrugated structures of graphene monolayer[9-13]. The observed peaks in this work are attributed to the emergence of quasi-localized quantum-well-like states confined in the 1D ripple of bilayer graphene. The other feature is the appearance of non-zero tunnelling conductance within the bandgap of the adjacent bilayer domains, which is consistent with the existence of topological edge states in the corrugation-induced stacking soliton[18,19]. Under a uniform external electric field, the stacking soliton between AB- and BA-stacked domains, to some extent, is equivalent to the domain wall separating two oppositely biased bilayer domains without stacking default[14-22].

Theories[18,19] and experiments[14,22] have found that a layer stacking domain wall can host valley-projected chiral edge modes in the presence of uniform electric field, which should apply to the observed corrugation induced stacking soliton. To further understand the observed spectra around the ripple, we calculate the electronic structure and the local density-of-state (LDOS) of a corrugation-induced stacking soliton separating the AB- and BA-stacked bilayer graphene domains, as shown in Fig. 2b and 2c (see methods and Supplemental Material for details of our calculation). In the calculation, we consider a tight-binding Hamiltonian with nearest-neighbour intra- and interlayer hopping for a ripple with the same width and curvature as the actual ripple shown in Fig. 1c. The topological edge states emerge in the stacking soliton of bilayer graphene (Fig. 2b) and the calculated LDOS (Fig. 2c) reproduce the main features of the spectra observed in the experiment. According to our calculation, we find that the curvature has no significant effect on the calculated electronic structure. Here we should point out that the topological edge states are absent if both the adjacent domains separated by the stacking soliton are AB-stacked (or BA-Stacked) bilayer regions (see Figure S6 of Supplemental Material).

We further explore the topological conducting channels along the corrugation induced stacking soliton using energy-fixed STS mapping, which can reflect the energy-resolved LDOS in real space. At the energies within the band gap of the adjacent AB and BA domains, clearly 1D conducting channels can be clearly observed along the stacking soliton, as shown in Fig. 3a. In our experiment, the LDOS at the ripple are almost energy independent, as long as the energy is within the band gap (see Figure S7 of Supplemental Material for more experimental data). Such a feature reflects the 1D nature of the topological conducting channels. In Fig. 3b, we show the recorded LDOS of the same area at energies away from the band gap, which becomes nearly homogeneous as expected. To compare our experimental data with theoretical predictions for the topological edge states, we calculate the spatial distribution of the gapless states around the corrugation-induced stacking soliton of the bilayer graphene. As shown in Fig. 3c and 3d, our calculation reveals that the topological edge states are mainly located at the ripple, matching well with the experimental data in Fig. 3a.

Figure 4a shows atomic resolution STS mapping of the topological edge states around the stacking soliton. The STS maps probe predominantly the LDOS of the top graphene layer. It is interesting to note that the amplitude of the wave function of the edge states in the AB-stacked (or BA-stacked) domain adjacent to the stacking soliton is mainly located on the *A* (or *B*) sublattice of top graphene layer. Therefore, the map exhibits contrasting triangular lattices, indicating the pronounced sublattice asymmetry of the edge states in both the AB- and BA-stacked domains while showing a hexagonal-like lattice at the centre of the stacking soliton (Fig. 4a). To compare with the experimental result, we plot a theoretical distribution of the topological gapless edge states in sublattices of the top graphene layer in Fig. 4b. Obviously, our experimental observations are well reproduced by the theoretical calculations. This further confirms that the observed 1D conducting channel along the ripple of the bilayer graphene arises from the valley-projected topological gapless modes due to the change in stacking order of biased bilayer graphene.

Besides the hexagonal-like lattice, the atomic-resolved STS maps of the topological edge states in the stacking soliton also exhibit another important feature: there is a larger honeycomb pattern, forming an $\sqrt{3}\times\sqrt{3}R30°$ unit cell (red diamond), with a darker depression at every other graphene hollow position (Fig. 4a). The emergence of the $\sqrt{3}\times\sqrt{3}R30°$ interference is known as a clear evidence of valley mixing (or intervalley scattering) in graphene[30,31]. The existence of intervalley scattering in the stacking soliton is further confirmed by fast Fourier transforms (FFT) of the STS maps such as the one shown in Fig. 4a (see Figure S8 of Supplemental Material for more experimental data). The FFT image, as shown in Fig. 4c, shows two sets of periodicities in the system with the outer six peaks corresponding to the graphene reciprocal lattice. The six inner peaks do not occur in pristine graphene and correspond to the intervalley scattering. Here, we point out that the intervalley scattering likely arises from the in-plane strain in the nanoscale ripple, visible in Fig. 4a, instead of any atomic defect that is clearly absent in the ripple. Such a result indicates that valley mixing of the valley-projected edge states, mainly located at the stacking soliton with the width of the order of 10 nm, is inevitable and may even be engineered by controlling the corrugation.

In the absence of intervalley scattering, electrons of edge states flowing in one direction cannot be scattered backward because the backward channels exist only in the other valley, as shown in Fig. 4d. The valley-projected edge states of the stacking soliton are analogous to the helical edge states of quantum spin Hall insulators[32-35]. In the former case there is a valley-momentum locking while a spin-momentum locking in the latter case. According to the electronic structure of the stacking soliton in bilayer graphene (Fig. 2b), it is clear that *K* and *K'* valleys each have two topological edge states per spin. Therefore, this 1D valley-projected conducting channel should exhibit quantized conductance of $4e^2/h$ in the absence of valley mixing. However, the back scattering between the counter-propagating valley-projected edge channels is not completely prohibited by any symmetry and is thus inevitable in the presence of lattice-scale defects or deformation. Electron-electron interactions and disorders may also enhance the possible pair backscattering[36]. Nevertheless, the observed valley mixing in the nanoscale ripple can dramatically affect the transport properties and, in particular, reduce the conductance along the stacking soliton[18]. This fact may explain the detected lower conductance, $< 4e^2/h$, along domain walls of gapped graphene bilayers in recent experiments[14,15].

In conclusion, we demonstrate that corrugations in bilayer graphene can generate incommensurate stacking solitons, hosting topological edge states, between commensurate AB-stacked and BA-stacked domains. Our results provide a new way to generate and tune stacking solitons, which are crucial for exploring exotic valley physics and exotic topological phases in bilayer graphene[20,37].

**Acknowledgments**

This work was supported by the National Basic Research Program of China (Grants Nos. 2014CB920903, 2013CBA01603, 2014CB920901), the National Natural Science Foundation of China (Grant Nos. 11674029, 11422430, 11374035, 11374219,11504008), the program for New Century Excellent Talents in University of the Ministry of Education of China (Grant No. NCET-13-0054). L.H. also acknowledges support from the National Program for Support of Top-notch Young Professionals. F.Z. is supported by UT Dallas research enhancement funds.

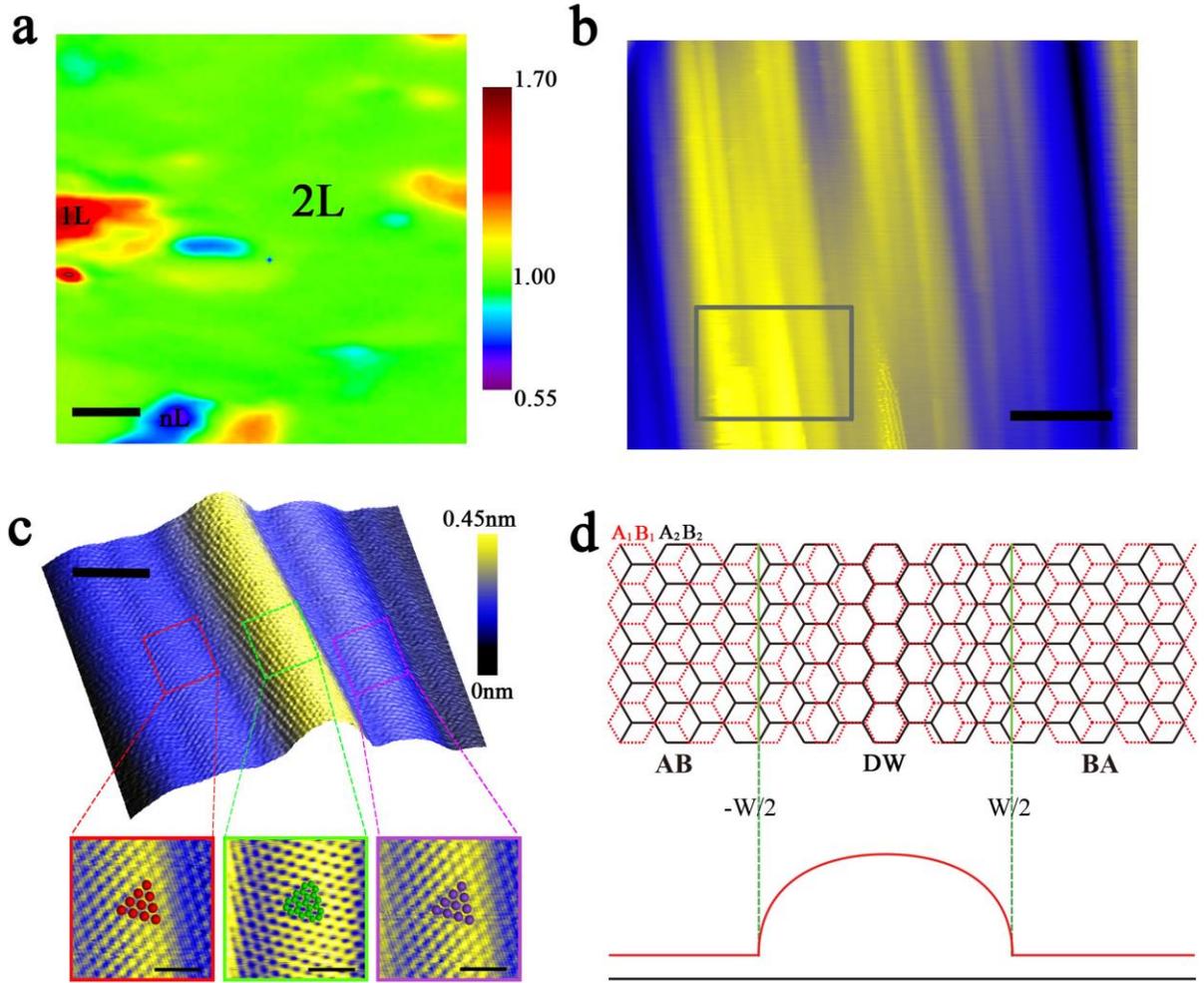

**Figure 1.** Corrugation induced stacking soliton in bilayer graphene. **a.** A 30 μm×30 μm $I_{2D}/I_G$ Raman mapping of the sample grown on a Rh foil showing that the region is mainly covered with bilayer graphene. Scale bar: 5 μm. **b.** A 50 nm ×40 nm STM image in the bilayer graphene region showing continuous corrugations with different widths and bending degree ($V_{sample}$ = -300 mV and $I$ = 0.4 nA). Scale bar: 10 nm. **c.** The 15 nm ×12 nm STM image in 3D view of a typical corrugation taken from the region marked by grey square in **a** ($V_{sample}$ = 180 mV and $I$ = 13 pA). Scale bar: 3 nm. The 3 nm ×3 nm atomic resolution STM images recorded aside the corrugation show triangular lattices, whereas the STM image recorded in the corrugation exhibits hexagonal lattices. Scale bars: 1 nm. **d.** A schematic image of a corrugation induced stacking soliton in bilayer graphene. The corrugated region (DW) with width of $W$ separates the AB-stacked and BA-stacked bilayer graphene domains.

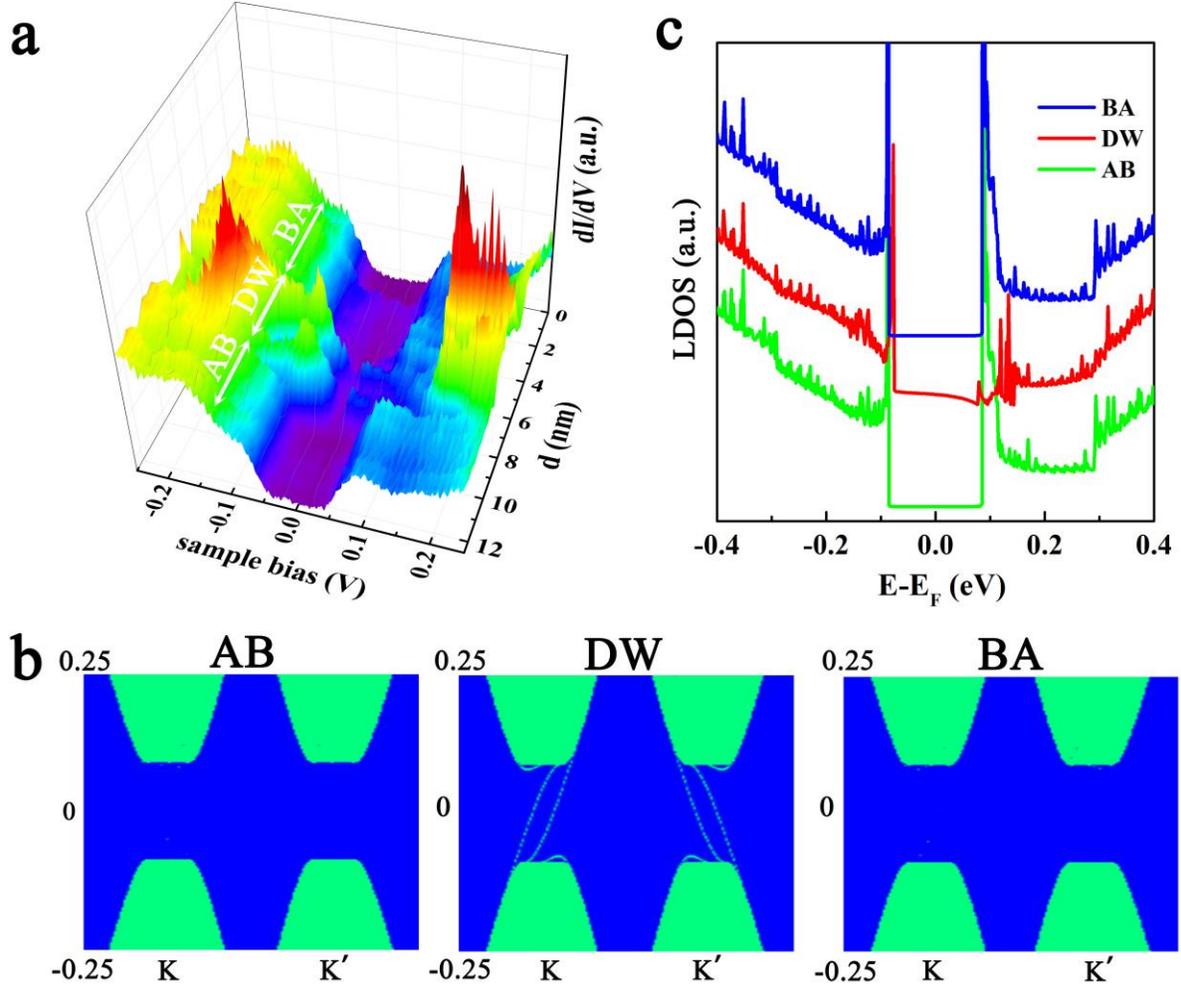

**Figure 2.** Electronic structures of the corrugation induced stacking soliton. **a.** STS spectra map taken across the stacking soliton from AB- to BA-stacked bilayer region. **b.** The projected LDOS as functions of momentum *k* and energy *E* in the three regions, *i.e.*, the AB-stacked bilayer, the corrugation induced stacking soliton (DW) and the BA-stacked bilayer. **c.** Calculated DOS spectra for AB, BA and DW regions. Because of computational limitations, the calculated systems are too small to avoid the emergence of van Hove singularities in both conduction and valence bands, as shown in panel c, due to quantum confinement.

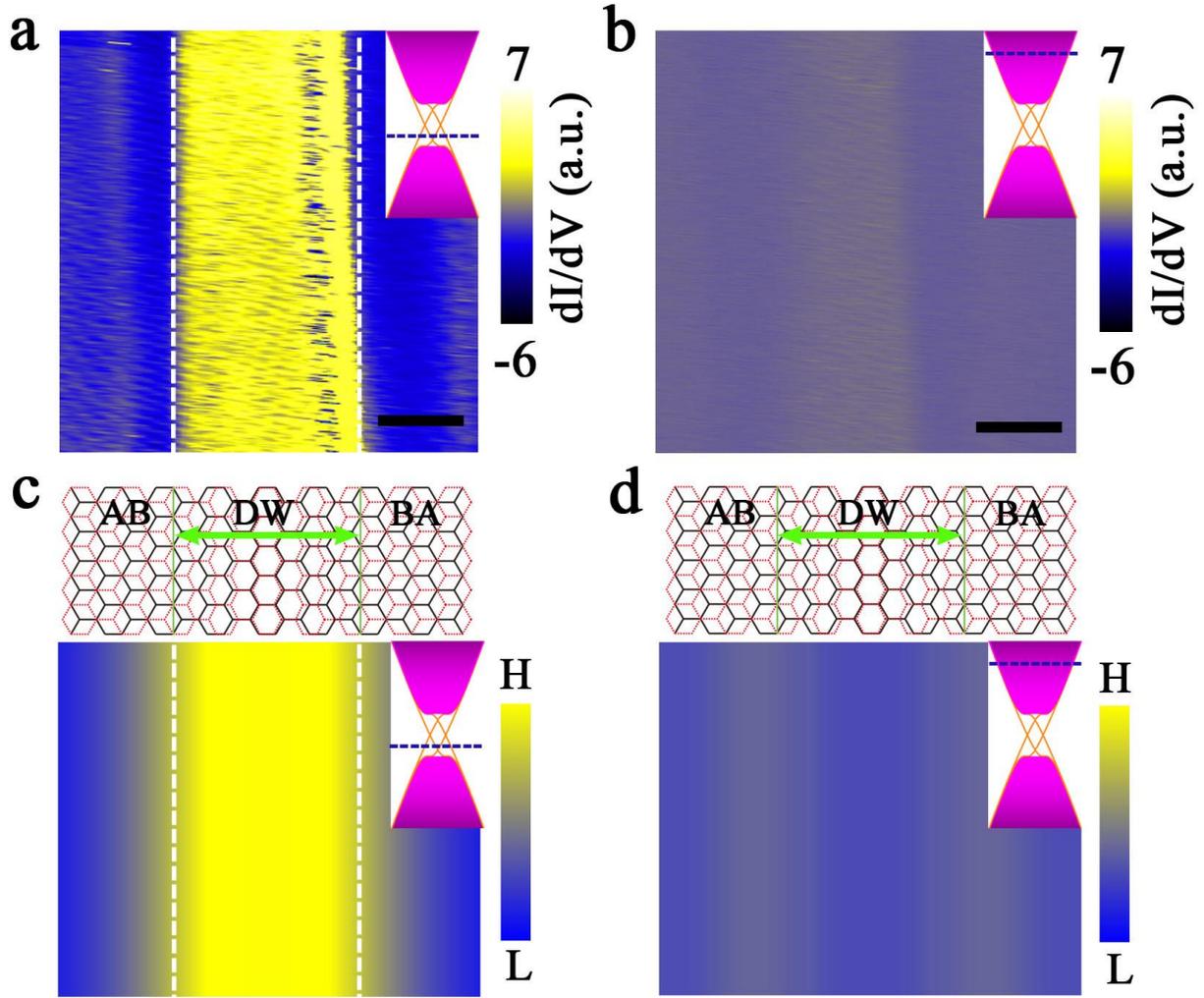

**Figure 3.** 1D conducting channel along the corrugation induced stacking soliton. **a, b.** STS maps recorded along the stacking soliton with the fixed sample bias of -15 mV in **a** and 300 mV in **b**, respectively. The edges of the stacking soliton are marked by two white dotted lines. The insets show the schematic band structure of the stacking soliton, in which the blue dotted lines mark the energies where each STS mapping is taken. Scale bars: 2 nm. **c, d.** Calculated spatial distributions of the localized state around the stacking soliton with the energies of -15 meV and 300 meV, respectively.

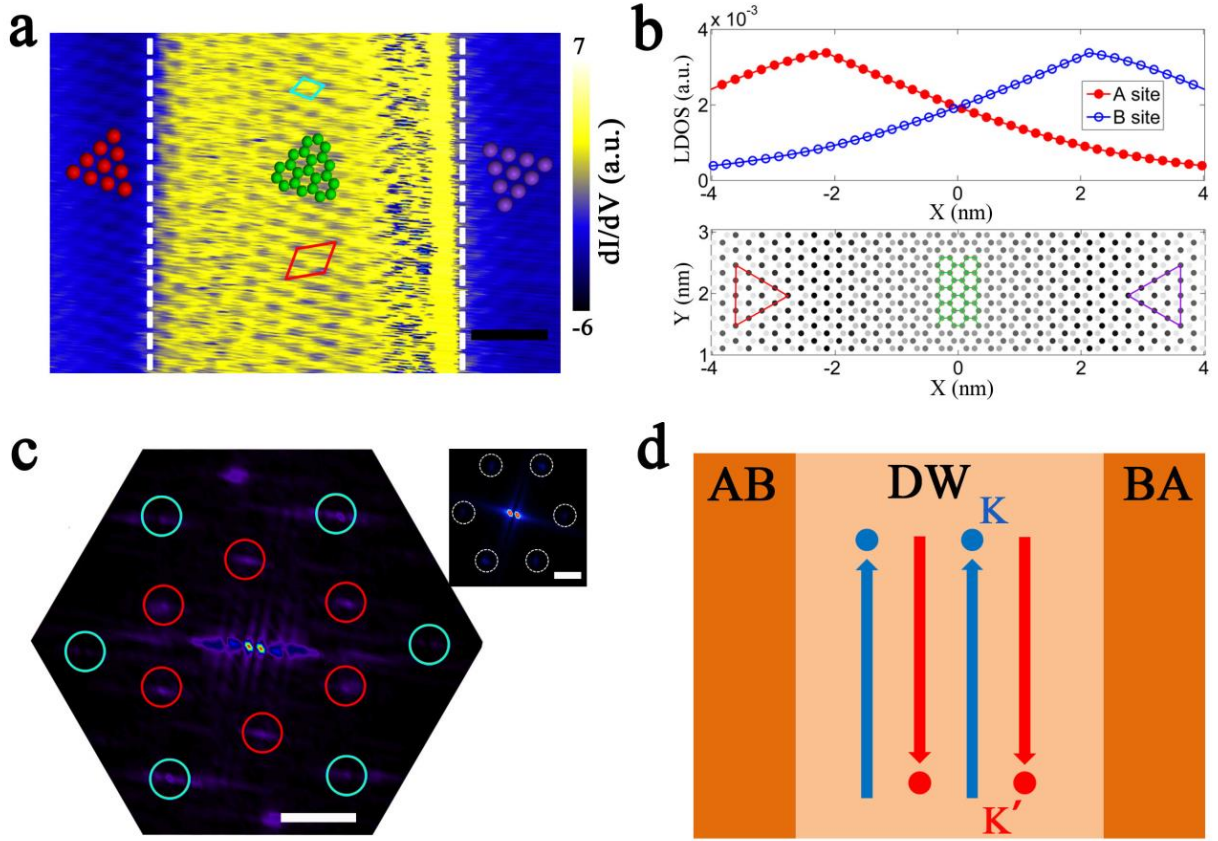

**Figure 4.** Corrugation induced intervalley scattering at the stacking soliton. **a.** A 6.5 nm ×4.5 nm atomic-resolved STS map recorded along the corrugation induced stacking soliton with the sample bias of 15 mV. Scale bar: 1 nm. The green dots show the hexagonal lattices in the center of the stacking soliton, while the red and purple dots present the triangular lattices observed aside the corrugation. The red diamond reflects the intervally-scattering supercell and the blue diamond is the graphene unit cell. **b.** The calculated LDOS (upper) and spatial distribution (lower) of topological edge states for the top-layer A and B sublattices around the stacking soliton. **c.** FFT of the atomic resolution STS map in panel **a**. The six outer peaks (blue circles) are the graphene Bragg peaks, and the six inner peaks (red circles) represent the intervalley scattering in reciprocal space. The inset shows the FFT of an atomic resolution STM image of graphene with the six spots corresponding to the reciprocal lattice of graphene. Scale bars: 15 Gm$^{-1}$. **d.** The schematic image of topological edge states counter-propagating at the $K$ and $K'$ valleys along a stacking soliton (DW) of gapped graphene bilayer.